\documentclass[prl,superscriptaddress,twocolumn,floatfix,linenumbers]{revtex4}
\usepackage{mathrsfs}
\usepackage{amsmath,amsfonts,amssymb, latexsym}
\usepackage{graphicx}% Include figure files
\usepackage{dcolumn}% Align table columns on decimal point
\usepackage[usenames]{color}
\usepackage{bm}% bold math
\usepackage{subfigure}

\definecolor{BrickRed}{cmyk}{0,0.89,0.94,0.28}%%%PANTONE 1805
\definecolor{MidnightBlue}{cmyk}{0.98,0.13,0,0.43}%%%PANTONE 302
\definecolor{DarkGreen}{rgb}{0,0.7,0.1}

\definecolor{RedViolet}{cmyk}{0.07,0.90,0,0.34}%%%PANTONE 234
\definecolor{SeaGreen}{cmyk}{0.69,0,0.50,0}%%%PANTONE 3268
\definecolor{FireOrange}{rgb}{1.,.294,.247}

\begin{document}
\title{Dirac semimetal in three dimensions}
\author{S. M. Young}
\affiliation{The Makineni Theoretical Laboratories, Department of Chemistry, University of Pennsylvania, Philadelphia, PA 19104-6323, USA}
\author{S. Zaheer}
\affiliation{Department of Physics and Astronomy, University of Pennsylvania, Philadelphia, PA 19104-6396, USA}
\author{J. C. Y. Teo\footnote{Present address: Department of Physics, University of Illinois at Urbana-Champaign, Urbana, Illinois 61801-3080, USA}}
\affiliation{Department of Physics and Astronomy, University of Pennsylvania, Philadelphia, PA 19104-6396, USA}
\author{C. L. Kane}
\affiliation{Department of Physics and Astronomy, University of Pennsylvania, Philadelphia, PA 19104-6396, USA}
\author{E. J. Mele}
\affiliation{Department of Physics and Astronomy, University of Pennsylvania, Philadelphia, PA 19104-6396, USA}
\author{A. M. Rappe}
\affiliation{The Makineni Theoretical Laboratories, Department of Chemistry, University of Pennsylvania, Philadelphia, PA 19104-6323, USA}
\date{\today}
\begin{abstract}
We show that the pseudo-relativistic physics of graphene near the Fermi level can be extended to three dimensional (3D) materials. Unlike in phase transitions from inversion symmetric topological to normal insulators, we show that particular space-groups also allow 3D Dirac points as symmetry protected degeneracies. We provide criteria necessary to identify these groups and, as an example, present \emph{ab initio} calculations of $\beta$-cristobalite ${\rm BiO_2}$ which exhibits three Dirac points at the Fermi level. We find that $\beta$-cristobalite ${\rm BiO_2}$ is metastable, so it can be physically realized as a 3D analog to graphene.
\end{abstract}
\maketitle
\begin{figure}[t]
{
\subfigure[]{ \includegraphics [width=1in]{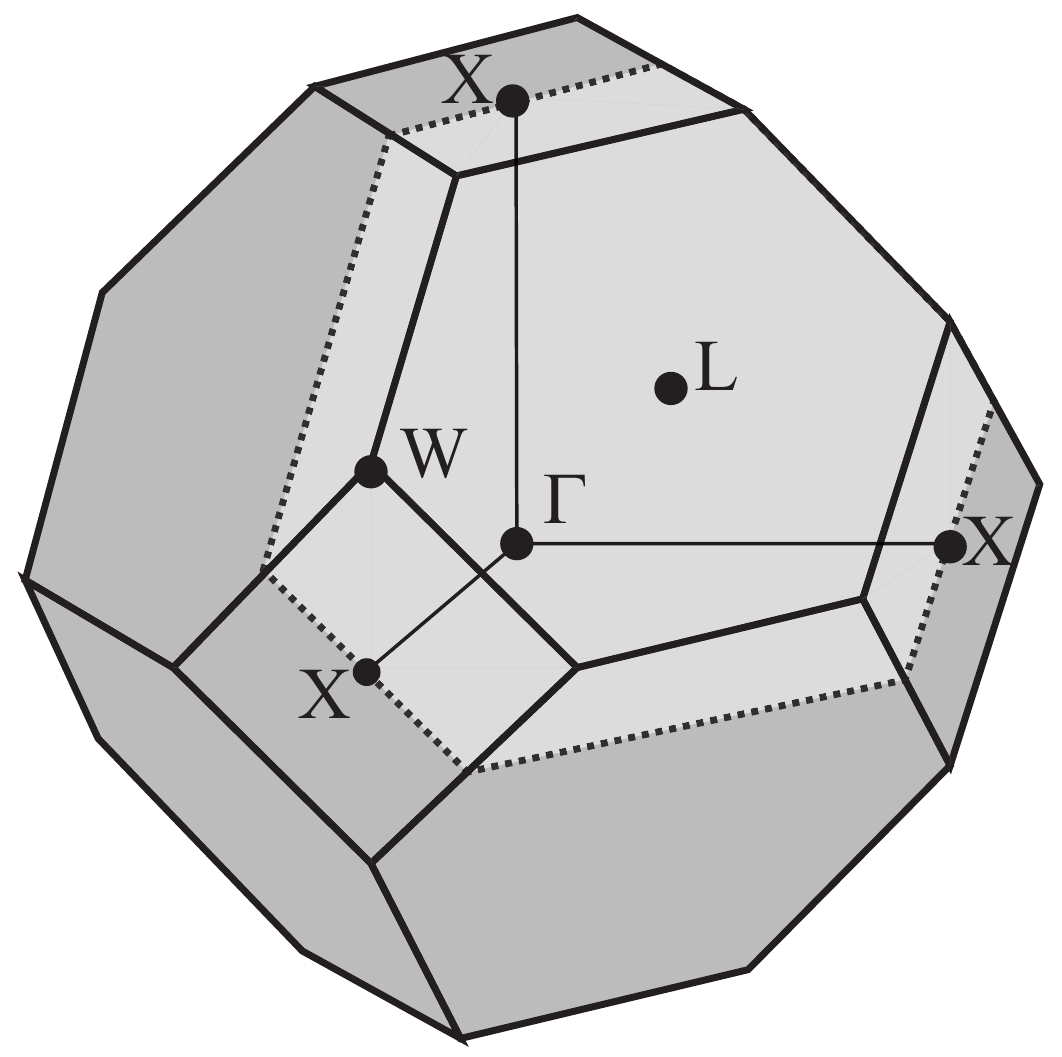}\label{fig:fcc}}
\subfigure[]{ \includegraphics [width=2in]{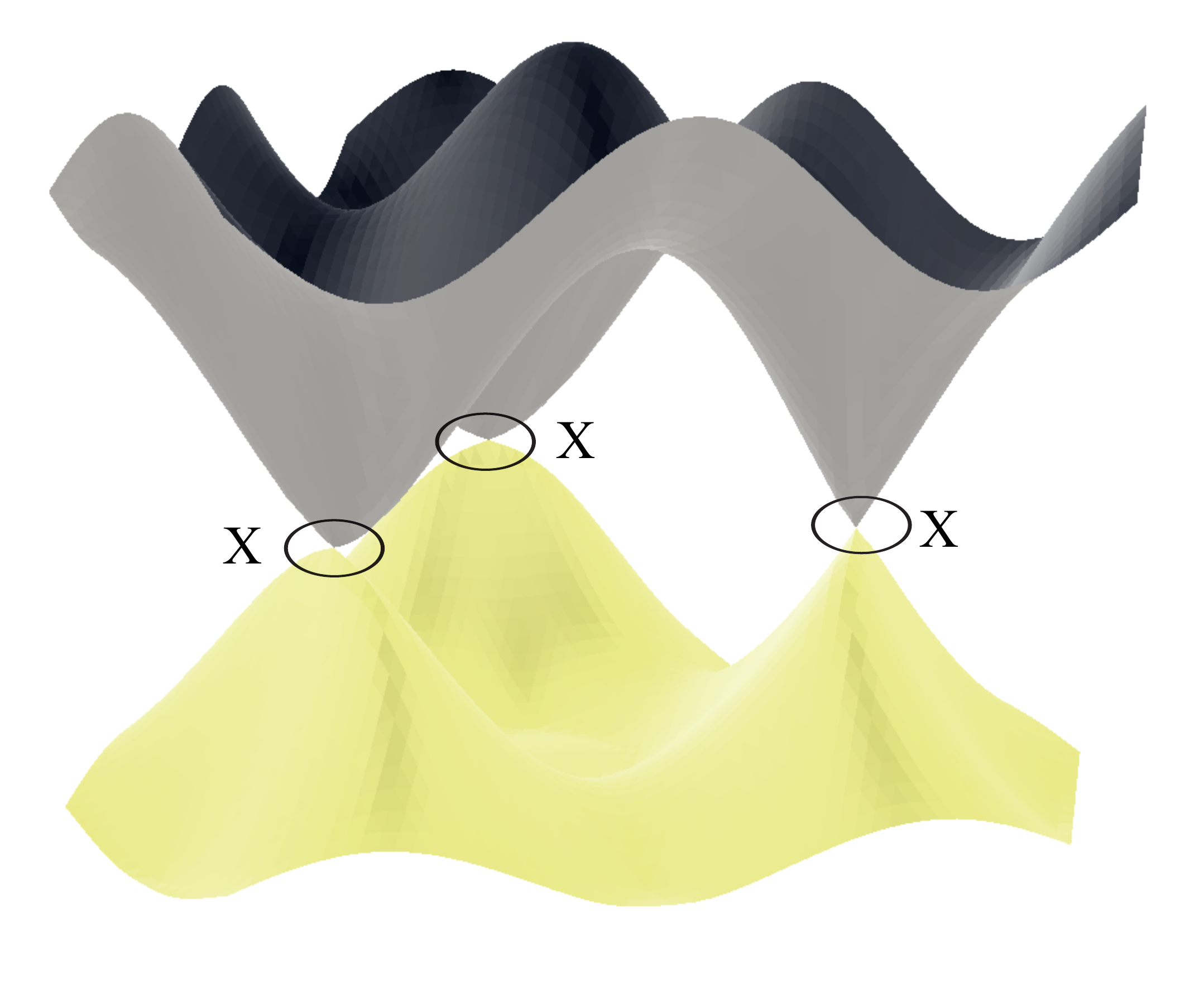}\label{fig:3dimage}}
\subfigure[]{ \includegraphics [width=3in]{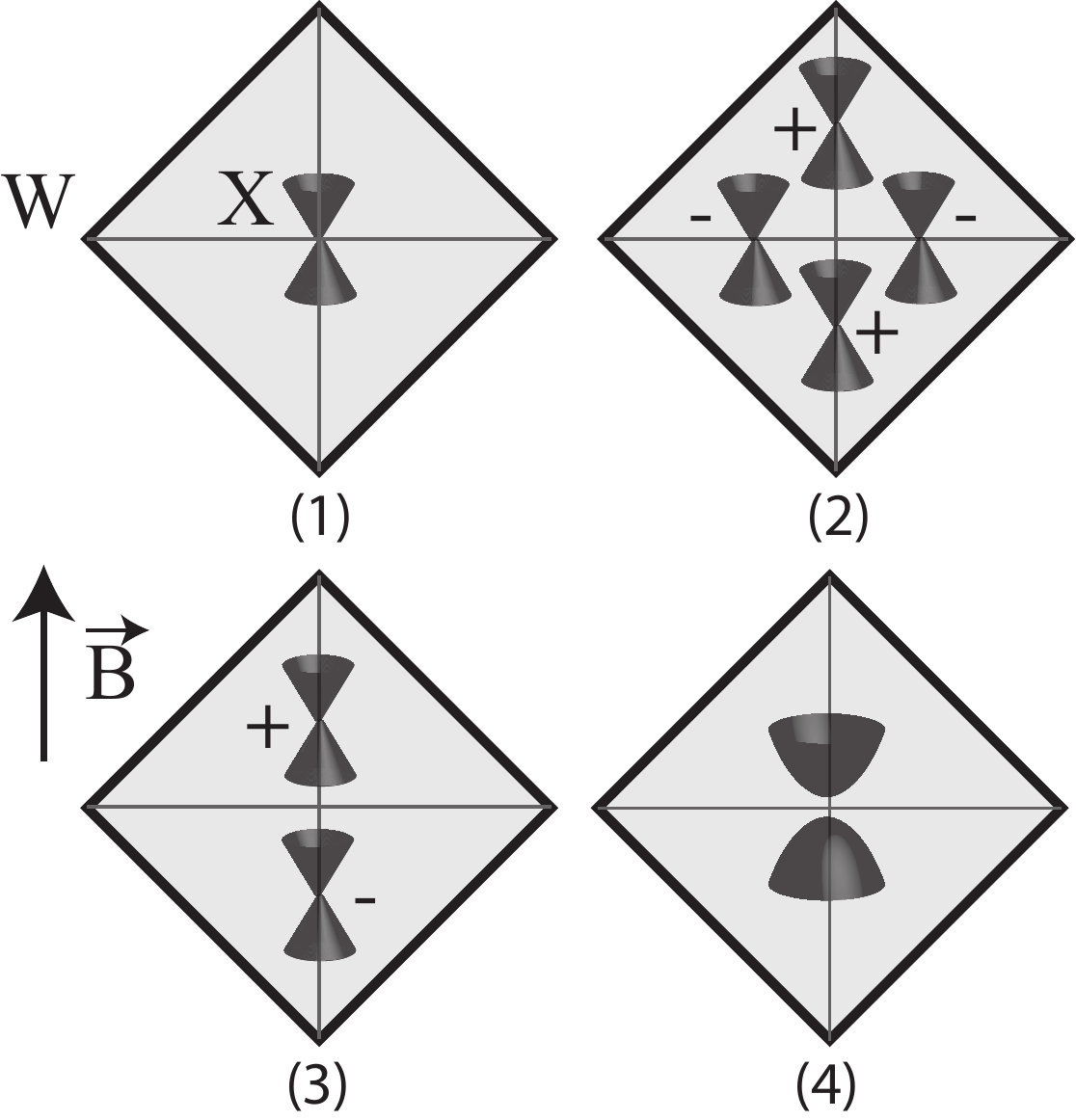}\label{fig:weyl}}
}
\caption{{3D Dirac semimetal in $\beta$-cristobalite ${\rm BiO_2}$.~\subref{fig:fcc} Brillouin zone (BZ) of the FCC lattice. The plane highlighted in gray joins the three symmetry related $X$ points. Other high symmetry points are also indicated.~\subref{fig:3dimage} Conduction and valence bands of $\beta$-cristobalite ${\rm BiO_2}$ are plotted as functions of momentum on the plane highlighted in gray on the left. Each band is two-fold degenerate due to inversion symmetry. Dirac points appear at the center of the three zone faces of the BZ.~\subref{fig:weyl} Dirac, Weyl and insulating phases in the diamond lattice. (1) The states at the Dirac point at $X$ span a four dimensional projective representation of the little group at $X$ which contains a four-fold rotation accompanied by a sub-lattice exchange operation. (2) Four Weyl points on the zone face due to a small inversion breaking perturbation. The Chern number of each Weyl point is indicated. (3) Two Weyl points appear on the line from $X$ to $W$ for a T-breaking Zeeman field ${\bf B}$ oriented along that direction. ${\bf B}$ oriented along other directions gaps all the Dirac points by breaking enough rotational symmetry that no two-dimensional representations are allowed. (4) Gapped phase obtained by breaking the four-fold rotation symmetry or by applying a magnetic field in any direction except along $\hat{x}$, $\hat{y}$, or $\hat{z}$. The insulating phase can be a normal, strong or a weak topological insulator~\cite{Fu07p106803}.}}
\end{figure}
In a Dirac semimetal, the conduction and valence bands contact only at discrete (Dirac) points in the Brillouin zone (BZ) and disperse linearly in all directions around these critical points. In two dimensions, spinless graphene exhibits such point-like degeneracies between the conduction and valence bands: the low energy effective theory at each of the critical points takes the Dirac form, $\hat{H}({\bf k}) = v(k_x \sigma_x +  k_y \sigma_y)$ where  $\vec{\sigma} = \{\sigma_x, \sigma_y, \sigma_z\}$ are the Pauli matrices and $v\not = 0$~\cite{CastroNeto09p109}. The existence of Dirac points near the Fermi level is responsible for many important properties of graphene such as high electron mobility and conductivity. However these Dirac points are not robust because they can be gapped by a perturbation proportional to $\sigma_z$. Spin-orbit coupling doubles the number of states and gaps the Dirac points~\cite{Kane05p226801}; however the splitting is very small because carbon is a light atom.  

In 3D, the analogous (and slightly generalized) Hamiltonian is $\hat H({\bf k}) = v_{ij} k_i \sigma_j$.  Provided ${\rm det}[v_{ij}] \ne 0$, $\hat H({\bf k})$ is robust against perturbations because it uses all three Pauli matrices. This Hamiltonian is called a Weyl Hamiltonian because it describes two linearly dispersing bands that are degenerate at a (Weyl) point. The robustness of a Weyl point can be quantified by the Chern number of the valence band on a sphere surrounding the point, which takes values ${\rm sgn}({\rm det}[v_{ij}]) = \pm 1$.   If a Weyl point occurs at some BZ momentum ${\bf k}$, time reversal (T) symmetry requires that another Weyl point occur at ${\bf -k}$ with equal Chern number.  However, the total Chern number associated with the entire Fermi surface must vanish.  Hence there must exist \emph{two more} Weyl points of opposite Chern number at ${\bf k'}$ and ${\bf -k'}$. Inversion (I) symmetry requires that Weyl points at ${\bf k}$ and ${\bf -k}$ have opposite Chern number. Hence under both T and I symmetries, ${\bf k} = {\bf k'}$ and the effective Hamiltonian involves four linearly dispersing bands around ${\bf k}$. Such a Hamiltonian is called a Dirac Hamiltonian, and it is not robust against perturbations because there are additional $4\times 4$ Dirac matrices that can be used to open a gap at the Dirac point.

The Fermi surface of a Dirac semimetal consists entirely of such point-like (Dirac) degeneracies. 3D Dirac semimetals are predicted to exist at the phase transition between a topological and a normal insulator when I-symmetry is preserved~\cite{Murakami07p356,Youngp11085106} (Ref.~\cite{Smithp056401} demonstrates such a Dirac point degenerate with massive bands.) If either I or T-symmetry is broken at the transition, a Dirac point separates into Weyl points and one obtains a Weyl semimetal (Fig.~\ref{fig:weyl}). The topological nature of Weyl points gives rise to interesting properties such as Fermi-arc surface states~\cite{Vishwanath11p205101} and pressure induced anomalous Hall effect~\cite{Yang11p075129}. Recent proposals to design a Weyl semimetal have been predicated upon the existence of a parent Dirac semimetal which splits into a Weyl semimetal by breaking I~\cite{Balents2011arXiv} or T-symmetry~\cite{Burkov11p127205}. Ref.~\cite{Manesp1109v2} demonstrates the existence of bulk chiral fermions due to crystal symmetry in single space-groups. 

\begin{figure}[t]
{
\subfigure[]{ \includegraphics [width=2in]{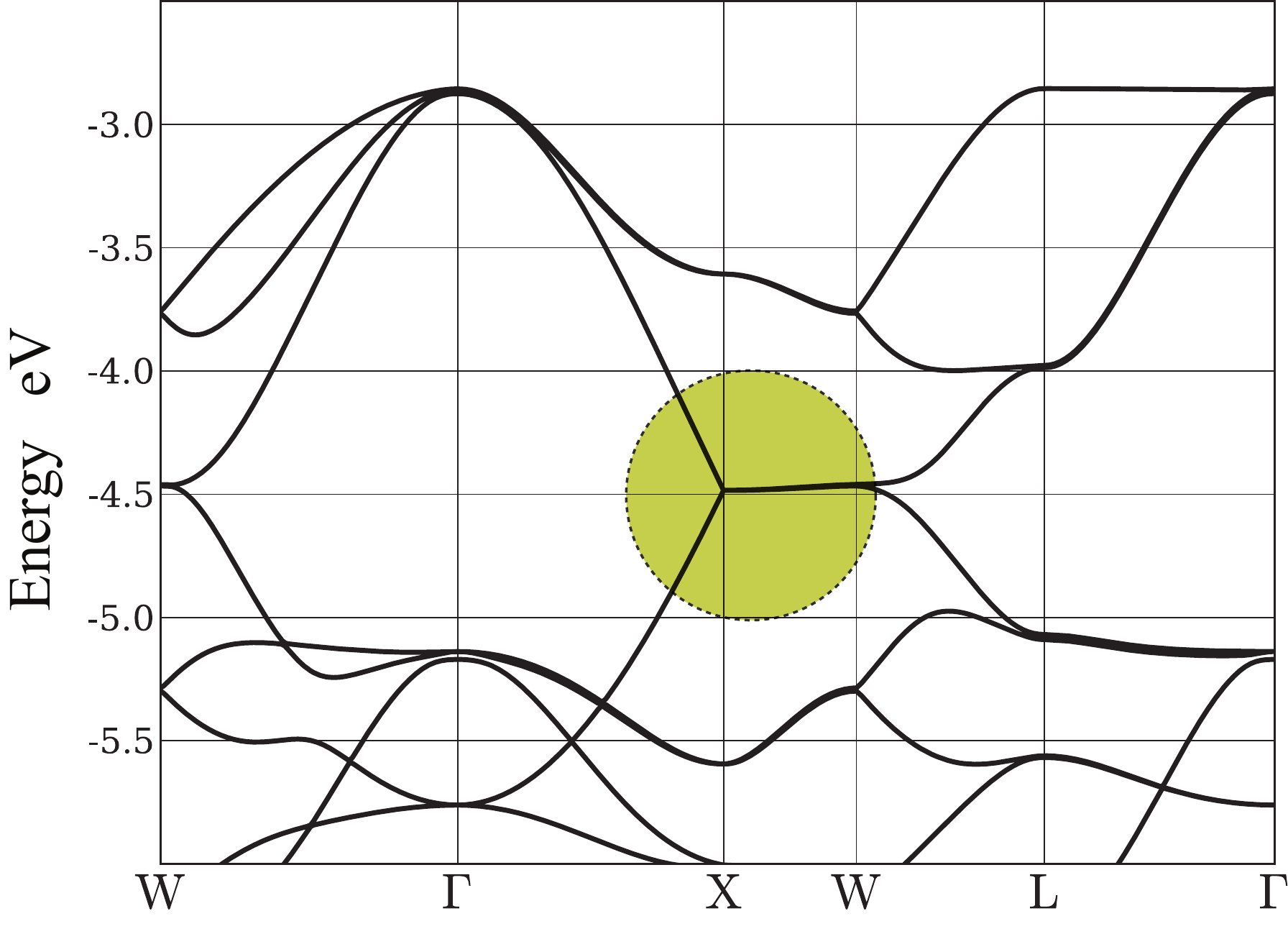}\label{fig:sio2}}
\subfigure[]{ \includegraphics [width=1in]{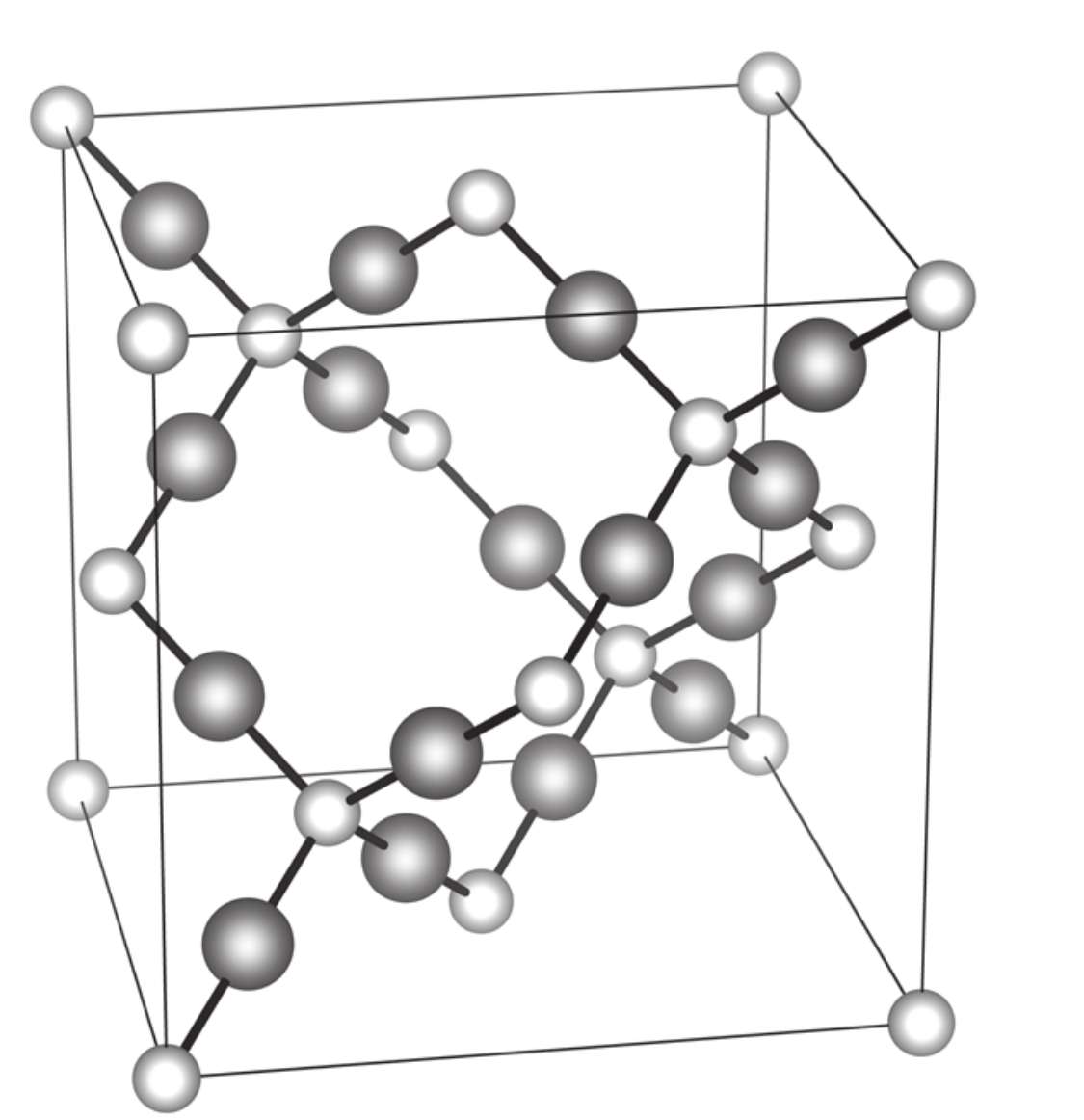}\label{fig:beta}}
}
\caption{ {~\subref{fig:sio2} Band structure of $\beta$-cristobalite $\rm SiO_2$. Energy bands are plotted relative to the Fermi level. Each band is two-fold degenerate due to inversion symmetry. The (highlighted) FDIR at $-4.5$~eV is split into two linearly dispersing bands between $X$ and $\Gamma$ while the two degenerate bands along $X$ and $W$ are weakly split. This FDIR is buried deep below the Fermi level.~\subref{fig:beta} The $\beta$-cristobalite structure of $\rm SiO_2$ ($\rm BiO_2$).  Silicon (bismuth) atoms (light gray) are arranged on a diamond lattice, with oxygen atoms (dark gray) sitting midway between pairs of silicon (bismuth).}}\label{struct}
\end{figure}

\begin{figure}[t]
{
\subfigure[]{ \includegraphics [width=1.5in]{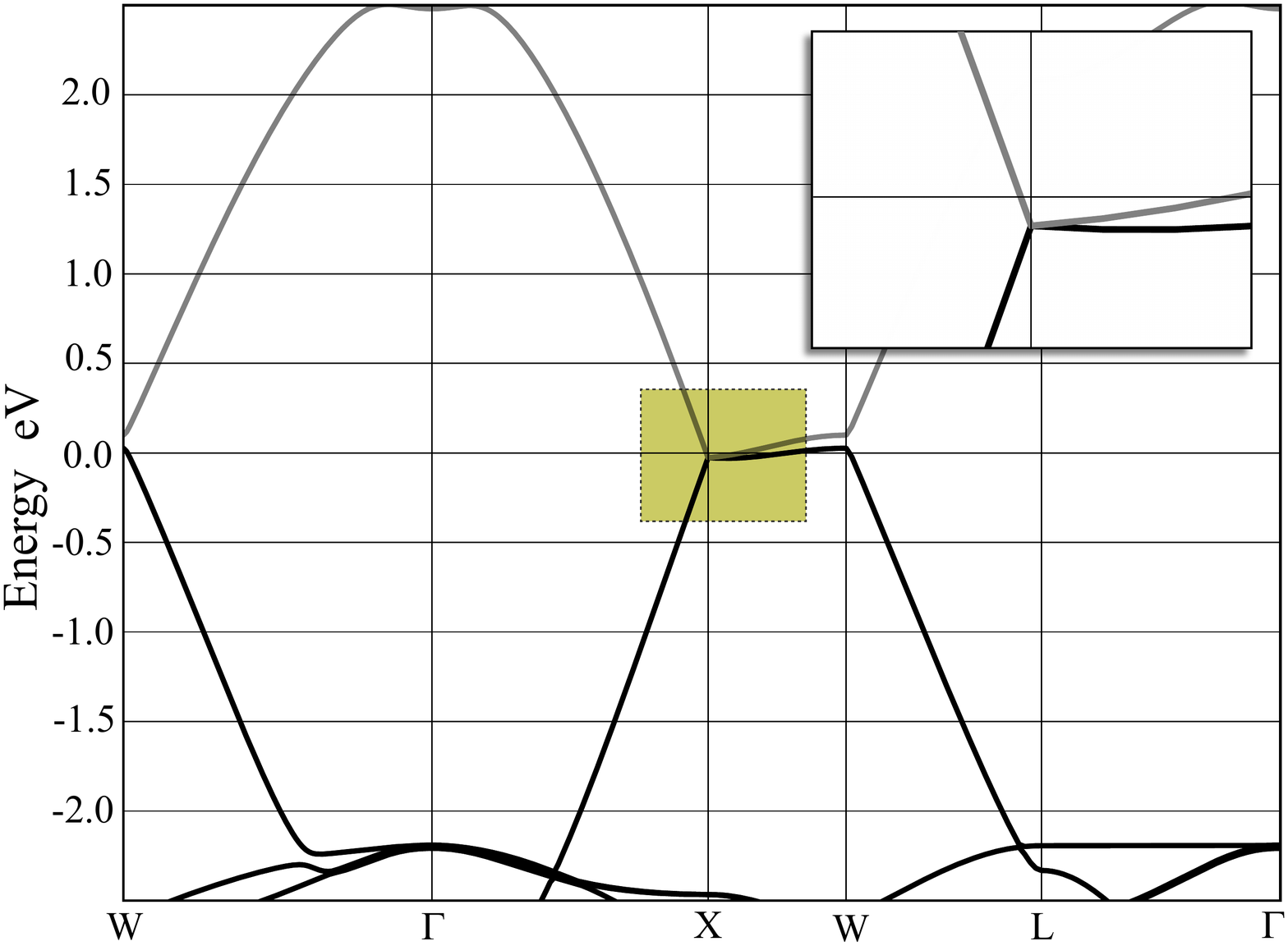}\label{fig:aso2}}
\subfigure[]{ \includegraphics [width=1.5in]{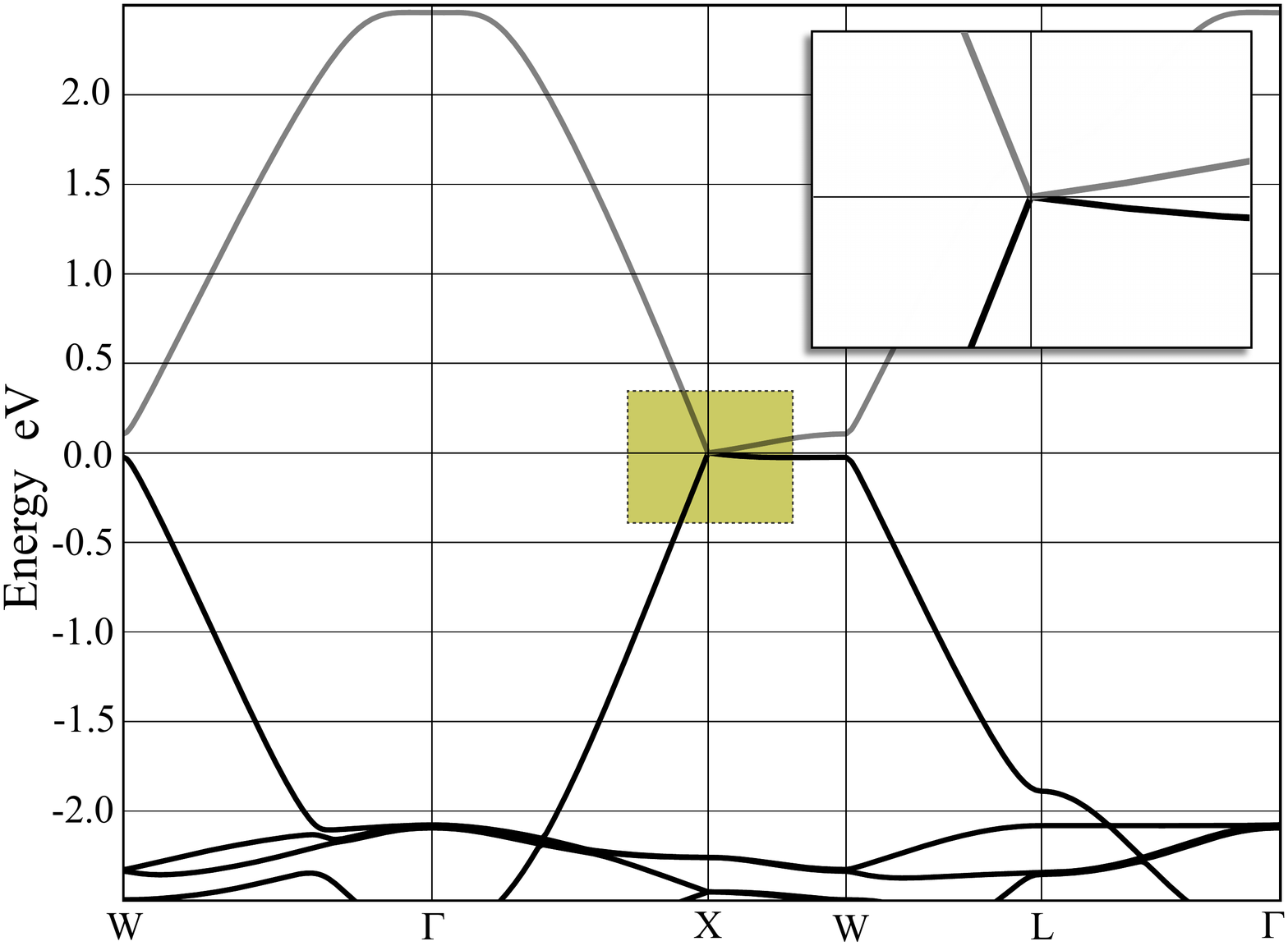}\label{fig:sbo2}}
\subfigure[]{ \includegraphics [width=1.5in]{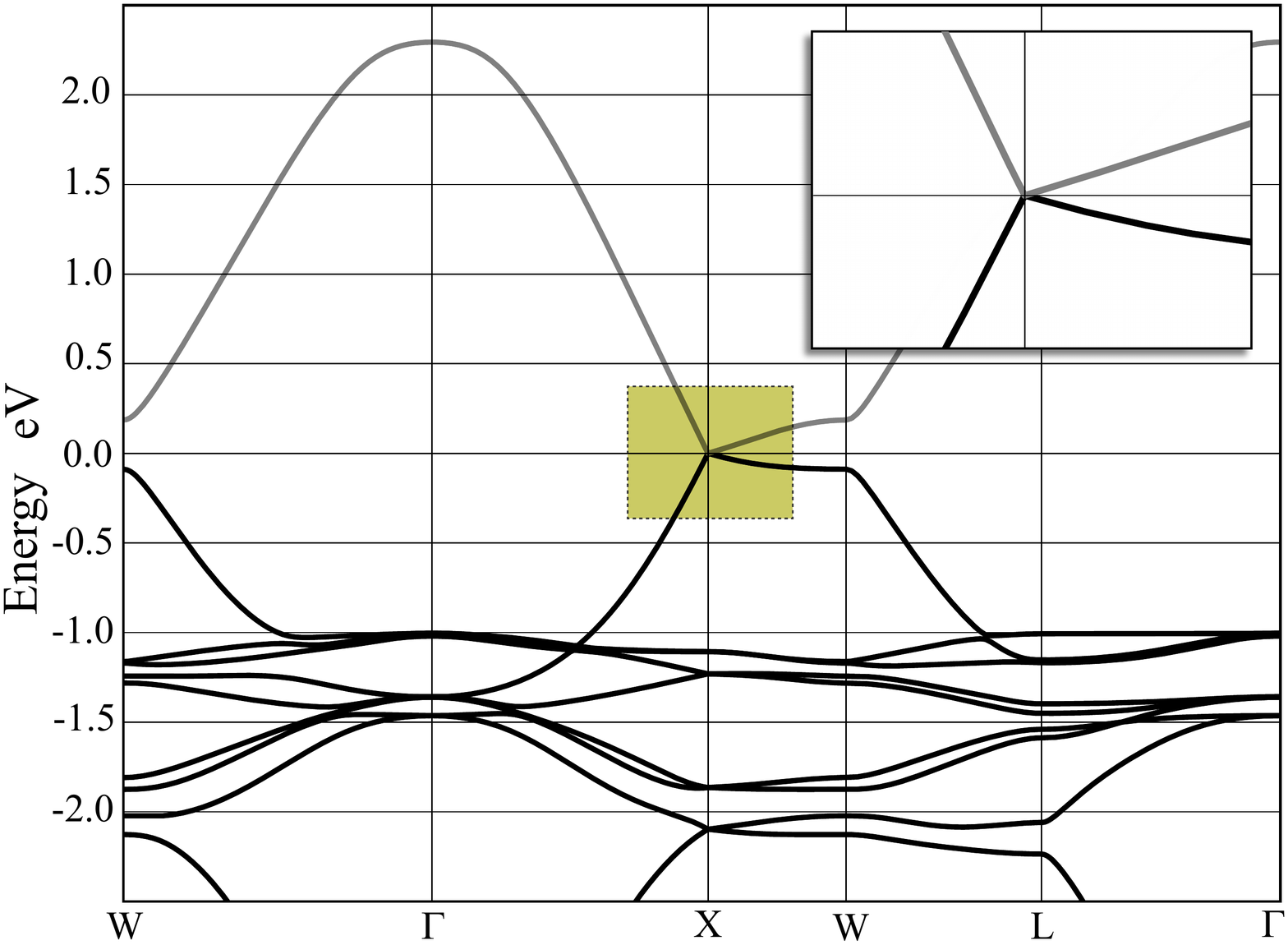}\label{fig:bio2}}
\subfigure[]{ \includegraphics [width=1.5in]{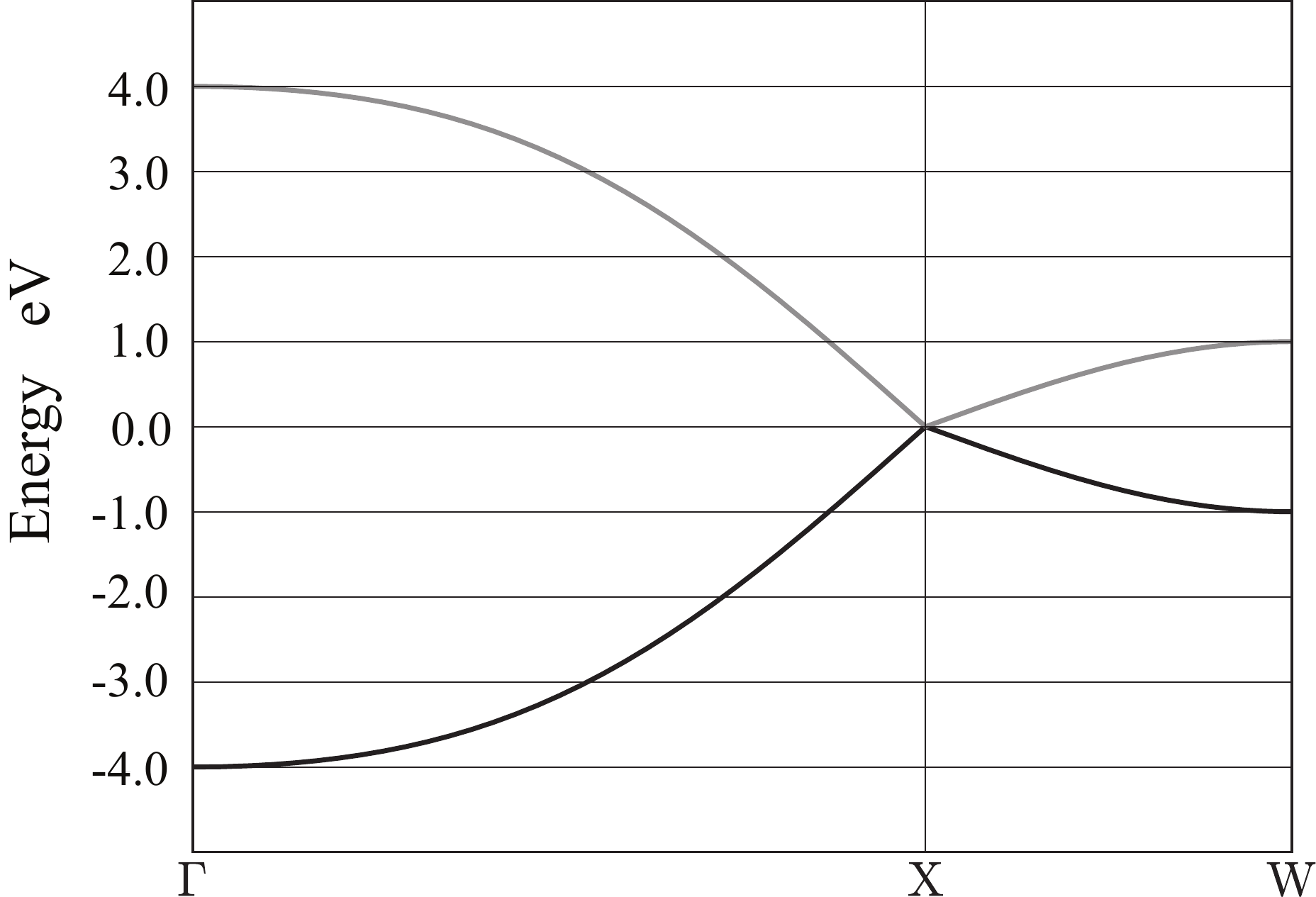}\label{fig:diamond}}
}
\caption{Band structures of~\subref{fig:aso2} AsO$_2$,~\subref{fig:sbo2} SbO$_2$, and~\subref{fig:bio2} BiO$_2$ in
the $\beta$-cristobalite structure, and~\subref{fig:diamond} $s$-states on a diamond lattice in the tight-binding model of Ref.~\cite{Fu07p106803}. Energy bands are plotted relative to the Fermi level. Each band is two-fold degenerate due to inversion symmetry. Insets: with
increasing atomic number of the cation, spin-orbit coupling widens the
gap along the line $V$ from $X$ to $W$. In $\rm BiO_2$ and $\rm SbO_2$, the dispersion around the $X$ point is linear in all directions indicating the existence of Dirac points at $X$. $\rm BiO_2$ and $\rm SbO_2$ are Dirac semimetals because their Fermi surface consists entirely of Dirac points.}\label{xo2}
\end{figure}

Dirac points that arise in a topological phase transition described above are accidental degeneracies. In general, two Weyl points with opposite Chern numbers annihilate each other unless their degeneracy is protected by additional space-group symmetry. Therefore we ask if a Dirac point can arise as a result of a crystallographic symmetry. Indeed certain double space-groups allow Dirac points at high symmetry points on the boundary of the BZ. As an example we present \emph{ab initio} calculations of $\beta$-cristobalite ${\rm BiO_2}$ (Fig.~\ref{fig:beta}) which exhibits Dirac points at three symmetry related $X$ points on the boundary of the FCC BZ (Figs.~\ref{fig:3dimage} and~\ref{fig:bio2}). This system realizes a Dirac degeneracy first encountered in a tight-binding model of $s$-states in diamond in Ref.~\cite{Fu07p106803} (the Fu-Kane-Mele model). In the absence of T-symmetry, two Weyl points with equal Chern numbers can be degenerate due to a point group symmetry as shown in Ref.~\cite{Fangp1111}. 

\begin{figure}[t]
\includegraphics[width=3in]{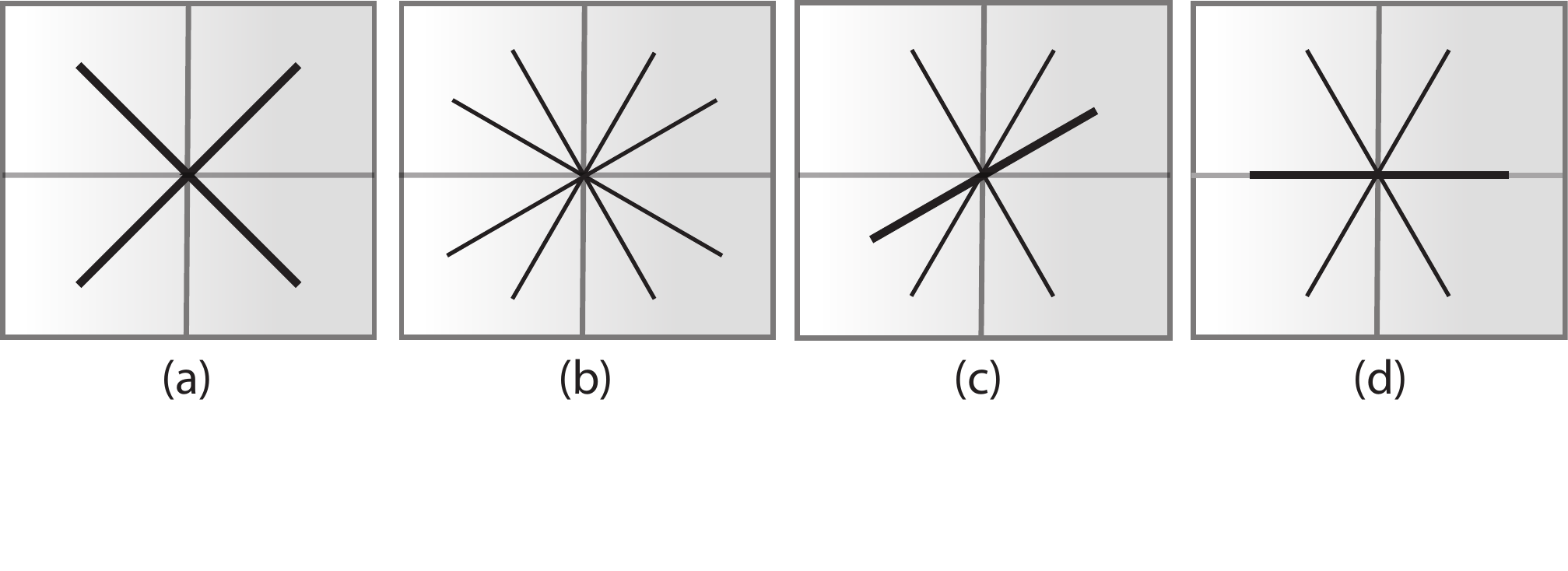}
\caption{Linear splitting of four-fold degenerate irreducible representations (FDIRs). If the symmetric kronecker product of an FDIR with itself contains the vector representation of the group to which the FDIR belongs, it will split in one of the four possible ways displayed above. (a) The FDIR splits into two two-fold degenerate bands. This situation is realized at the $X$ point of the FCC Brillouin zone in a diamond lattice. (b) The FDIR splits into four non-degenerate bands. This situation arises at the $\Gamma$ point in zincblende if mirror symmetry is broken (although the FDIR in zincblende develops a non-zero Chern number due to three-fold rotation symmetry at $\Gamma$). (c) The FDIR splits into two non-degenerate and one two-fold degenerate band with linear dispersion. (d) The splitting of the FDIR at $\Gamma$ in zincblende. The two-fold degenerate band is constrained to be flat implying quadratic dispersion along that direction. The Chern number of this representation is zero inspite of a three-fold rotation symmetry because the conduction and valence bands are degenerate away from $\Gamma$.}\label{foursplit}
\end{figure}

A 3D double space-group must satisfy the following criteria to allow a Dirac point. It must admit four dimensional irreducible representations (FDIRs) at some point ${\bf k}$ in the BZ such that the four bands degenerate at ${\bf k}$ disperse linearly in all directions around ${\bf k}$ and the two valence bands carry zero total Chern number. If the little group $G_{\bf k}$ at ${\bf k}$ contains a three-fold or a six-fold rotation symmetry and the valence and conduction bands around ${\bf k}$ are non-degenerate, the Chern number of the FDIR is guaranteed to be non-zero. This rules out symmorphic space-groups with FDIRs because they contain three-fold rotations. This also rules out interior BZ momenta because non-symmorphic little groups without three-fold rotations exhibit FDIRs only on the boundary of the BZ~\cite{Bradley72p1}. To guarantee linear dispersion of bands around ${\bf k}$, the symmetric kronecker product $[R_{\bf k} \times R_{\bf k}]$ of the FDIR with itself must contain the vector representation of $G_{\bf k}$~\cite{Hamermesh64p}. Finally, away from ${\bf k}$, the FDIR must split so that the valence and conduction bands are non-degenerate everywhere except at ${\bf k}$~(Fig.~\ref{foursplit}). 

We apply the above criteria to two important space-groups. The space-group of diamond (227, Fd3m), which is also the symmetry group of $\beta$-Cristobalite ${\rm BiO_2}$, exhibits FDIRs $R_{\Gamma}$ at $\Gamma$ and $R_{X}$ at $X$. $G_{\Gamma}$ contains three-fold rotation symmetry and $[R_{\Gamma} \times R_{\Gamma}]$ does not contain the vector representation of $G_{\Gamma}$. Therefore the $\Gamma$ point in a diamond lattice cannot host a Dirac point. $R_{X}$ is a projective representation of $G_{X}$ which does not have any three-fold rotations because all the point group operations in $G_{X}$ are those of the group $D_{4h}$. $[R_{X} \times R_{X}]$ contains the vector representation of $G_{X}$. Finally $R_{X}$ splits into either two doublets or four singlets away from $X$ (Figs.~\ref{foursplit}(a) and~\ref{foursplit}(b)). Therefore the $X$ point in space-group 227 is a candidate to host a Dirac semimetal if its FDIR can be elevated to the Fermi level. Indeed we show that $\beta$-Cristobalite ${\rm BiO_2}$ exhibits such a Dirac point at $X$, Fig.~\ref{fig:bio2}. The Dirac point at $X$ in the FKM model is also spanned by states belonging to $R_{X}$ (Fig.~\ref{fig:diamond}). 

The zincblende lattice (space-group 216, F$\bar{4}$3m) has an FDIR $R'_{\Gamma}$ at $\Gamma$ and the little group $G'_{\Gamma}$ has a three-fold rotation symmetry. $[R'_{\Gamma} \times R'_{\Gamma}]$ contains the vector representation of $G'_{\Gamma}$. Mirror symmetry in $G'_{\Gamma}$ requires $R'_{\Gamma}$ to split into a two-fold degenerate representation and two non-degenerate representations along the (111) axis, which is also the symmetry axis for the three-fold rotation. Time reversal symmetry requires that the two-fold degenerate band remain flat along the (111) axis, Fig.~\ref{foursplit}(d). Thus the lowest band carries Chern number 0, while the two flat bands carry 1 and -1. Therefore the dispersion of $R'_{\Gamma}$ is not Dirac-like along (111). 

In HgTe, which takes the zincblende lattice, the degenerate valence and conduction states at $\Gamma$ span $R'_{\Gamma}$ and constitute the entire Fermi surface. It is known that in HgTe the valence and conduction bands disperse linearly in two directions around $\Gamma$ and quadratically in a third (Fig.~\ref{foursplit}(d) and Ref.~\cite{Dresselhaus55p580}). One might ask if a perturbation might turn HgTe into a Dirac semimetal. However the zincblende lattice does not satisfy the criteria for 3D Dirac points as outlined above, so HgTe cannot host a Dirac semimetal. (a) $\Gamma$ is an interior point of the BZ and the little group at $\Gamma$ contains a three-fold rotation. (b) Mirror symmetry requires two bands to be degenerate along the (111) axis but since the Chern number must vanish, the degenerate bands must be flat and consist of a conduction and a valence band. This is why we see quadratic dispersion along the (111) axis. (c) Breaking mirror symmetry splits the degenerate flat band but then the Fermi surface develops other non-Dirac like pockets to compensate for the non-zero Chern number. (d) Breaking three-fold rotation symmetry splits the degeneracy at $\Gamma$ entirely and the material becomes a topological insulator~\cite{Konig07p766}.

We briefly discuss the theory behind the above criteria. We are interested in FDIRs of double space-groups at points ${\bf k}$ such that the valence and conduction bands are distinct in a small region around ${\bf k}$ and carry zero total Chern number. The Chern number of a degenerate representation can be determined up to an integer by the rotation eigenvalues of the valence bands. Electron states spanning an FDIR are equivalent to a $p_{\frac{3}{2}}$ quadruplet which exhibit eigenvalues $e^{\pm i3\pi/n}, e^{\pm i\pi/n}$ for a $2\pi/n$ rotation symmetry. Rotation eigenvalues of states at time reversed momenta about the degenerate point are complex conjugates. Therefore the FDIR will carry Chern numbers $\pm 1 \mod n$ for one valence band and $\pm 3 \mod n$ for the other with total Chern number $\pm 4 \mod n$ or $\pm 2 \mod n$ for the FDIR. This is zero only for $n=1,2,4$. If the conduction and valence bands are distinct in a small region around ${\bf k}$, the Chern number of the FDIR will be non-zero if the little group $G_{\bf k}$ contains a $2\pi/3$ or $2\pi/6$ rotation symmetry. In HgTe however, the little group at $\Gamma$ contains a three-fold rotation symmetry but the FDIR at $\Gamma$ has zero Chern number because one of the valence bands is degenerate with one of the conduction bands along the (111) axis. 

Non-symmorphic space-groups contain point group operations coupled with non-primitive lattice translations. For example, inversion interchanges the FCC sub-lattices in the diamond space-group.  Representations of non-symmorphic space-groups at momenta inside the BZ momenta are obtained from regular representations, while those at the \emph{surface} BZ momenta are obtained from \emph{projective} representations of the associated crystal point group. The factor system of the projective representation is chosen to implement the required non-primitive translation corresponding to the non-symmorphic point group operation~\cite{Bradley72p1}. A theorem by Schur guarantees that projective representations of a group can be obtained by restricting to the group elements the regular representations of a larger group called the central extension group~\cite{Bradley72p1}. The central extension of a group is obtained by taking its product with another finite Abelian group. The important point to emphasize is that representations of non-symmorphic space-groups are obtained from representations of central extensions of the 32 point groups. Central extension groups exhibit FDIRs even without three-fold rotations in the original point group. This is precisely why Dirac points can exist in 3D as symmetry allowed degeneracies.  

To realize a Dirac-like dispersion in the vicinity of an FDIR, some of the matrix elements ${\langle\psi_i| {\bf p}| \psi_j \rangle}$, where $|\psi_i\rangle$ span the FDIR, must be non-zero. This is guaranteed if the symmetric kronecker product of the FDIR with itself contains the vector representation of the central extension group to which the FDIR belongs~\cite{Hamermesh64p}. We restrict to the symmetric part of the kronecker product because matrix elements  ${\langle\psi_i| {\bf p}| \psi_j \rangle}$ correspond to level transitions between states spanning the same representation~\cite{Dresselhaus55p580}. Finally, the allowed representations in the vicinity of an FDIR should be such that each band disperses with non-zero slope in all directions. This is possible only if the valence band is distinct from the conduction band everywhere except at the Dirac point. Fig.~\ref{foursplit} illustrates the various possible ways in which an FDIR can split linearly. 

Although crystallographic symmetries determine whether 3D Dirac points can exist, physical and chemical considerations dictate whether they arise at the Fermi level without additional non-Dirac like pockets in the Fermi surface.  In the FKM model, the Dirac point at $X$ appears at the Fermi energy.  However, in known materials on a diamond lattice $s$-states appear below the Fermi energy. In realistic systems, additional orbitals hybridize with these $s$-states and bands cross the Fermi level at other points besides $X$. The problem is especially severe in space-group 227: without spin, the line $V$ from $X$ to $W$ is two-fold degenerate. With spin-orbit coupling, this line splits weakly for lighter atoms so the bands dispersing along this line can hybridize and introduce additional Fermi surface. Forcing species with $s^1$ valence states on the diamond lattice would fail to realize the FKM model. Indeed, {\it ab initio} calculations with group I elements and gold show that the splitting along $V$ is insufficient to overcome this dispersion. In some cases, additional bands crossed the Fermi level.  We performed \emph{ab initio} calculations using the plane wave density functional theory package quantum espresso~\cite{Giannozzi09p395502}, and designed non-local pseudopotentials~\cite{Rappe90p1227, Ramer99p12471} with spin-orbit interaction generated by OPIUM. 
 
We consider derivatives of the diamond lattice that remain in space-group 227. We place additional atoms in the lattice such that the configuration of added species allows its valence orbitals to either belong to the FDIR of interest, or appear away from the Fermi energy of the final structure. If the new species can split the nearby $p$ states of the existing atoms away from the $s$ levels, band crossing at the Fermi level can be avoided. 

One such structure is $\beta$-cristobalite SiO$_2$ (Fig.~\ref{fig:beta}), which consists of silicon atoms on a diamond lattice with oxygen atoms placed midway along each silicon-silicon bond~\cite{Calvert1991}. Oxygen atoms have two consequences: part of the O $p$-shell strongly hybridizes with the Si $p$-states, moving them away from the Si $s$-states, while the remaining O $p$-states span the same representation as the Si $s$-states. A Dirac point can be realized by Si $s-$O $p$ bonding/anti-bonding set of states. Fig.~\ref{fig:sio2} shows that the Si $s-$O $p$ bands are present and take a configuration similar to the valence and conduction bands in the FKM model, but appear well below the Fermi energy. Additionally, the bands are nearly degenerate along the line $V$ from $X$ to $W$ due to weak spin-orbit coupling.
 
Heavier atoms substituting Si both widen this gap and bring the FDIR of interest at $X$ to the Fermi level. Fig.~\ref{xo2} shows the band structures of compounds $\beta$-cristobalite $\rm XO_2$ where X = {As/Sb/Bi}. The change in chemical identity promotes the X $s-$O $p$ four-fold degeneracy at $X$ to the Fermi level, and stronger spin-orbit coupling widens the gap along $V$. $\rm BiO_2$ bears striking similarity to the FKM model, with linearly dispersing bands in a large energy range around a Dirac point at the Fermi level. Our calculations show that the phonon frequencies for $\beta$-cristobalite $\rm BiO_2$ at $\Gamma$ are positive, so it is a metastable structure. Further calculations reveal that it becomes unstable under uniform compression exceeding 2GPa, which represents a stability barrier of approximately 0.025eV per atom. On this basis, the possibility of synthesis appears promising. However, $\rm Bi_2O_4$ is also likely to take the cervantite structure (after $\rm Sb_2O_4$, which has similar stoichiometry~\cite{Thorntonp1271}) which is 0.5 eV per atom lower in energy as compared to $\beta$-cristobalite and 60$\%$ smaller in volume. Therefore we conclude that $\beta$-cristobalite $\rm BiO_2$ would be metastable if synthesized, although preventing it from directly forming the cervantite structure would be challenging. Nonetheless we have provided an existence proof of a Dirac semimetal in $\beta$-cristobalite $\rm BiO_2$ due to real atomic potentials at the DFT level. 

\emph{Acknowledgements.}
We thank Jay Kikkawa and Fan Zhang for helpful comments on the manuscript. This work was supported in part by the MRSEC program of the National Science Foundation under grant no. DMR11-20901 (S.M.Y.),  by the Department of Energy under grant no. FG02-ER45118 (E. J. M. and S. Z.), and by the National Science Foundation under grant nos. DMR11-24696 (A. M. R.) and DMR09-06175 (C. L. K. and J. C. Y. T.). S. M. Y. acknowledges computational support from the High Performance Computing Modernization Office.

\end{document}